# The twist-bend nematic phase of bent mesogenic dimer CB7CB and its mixtures

Michael R. Tuchband[a], Min Shuai[a], Keri A. Graber[a], Dong Chen[a], Leo Radzihovsky[a], Arthur Klittnick[a],

Lee Foley[b], Alyssa Scarbrough[b], Jan H. Porada[b], Mark Moran[b], Eva Korblova[b], David M. Walba[b],

Matthew A. Glaser[a], Joseph E. Maclennan[a], and Noel A. Clark [a*]

[a]Department of Physics and Soft Materials Research Center, University of Colorado,

Boulder, CO 80309-0390, USA

[b]Department of Chemistry and Biochemistry and Soft Materials Research Center, University of Colorado,

Boulder, CO 80309-0215, USA

*E-mail: noel.clark@colorado.edu

**Abstract**

Binary mixtures of the twist-bend nematic-forming liquid crystal CB7CB with the prototypical rod-like liquid crystal 5CB exhibit a twist-bend nematic ($N_{TB}$) phase with properties similar to those reported for neat CB7CB. The mixtures appear homogeneous, with no micron- or nano-scale segregation evident at any concentration. The linear dependence of the phase transition temperature on concentration indicates that these binary mixtures are nearly ideal. However, a decrease in the viscosity with the addition of 5CB allows the characteristic twist-bend stripe textures to relax into a state of uniform birefringence. We confirm the presence of nanoscale modulations of the molecular orientation in the mixtures by freeze-fracture transmission electron microscopy (FFTEM), further evidence of their twist-bend nature. We devise and implement a statistical approach to quantitatively measure the ground state pitch of the twist-bend phase and its mixtures using FFTEM. The addition of 5CB generally shifts the measured ground-state pitch distributions towards larger pitch. Interestingly, the pitch appears to increase discontinuously by ~10 nm at the 50 wt% concentration of 5CB, indicating that the twist-bend phase undergoes a structural transition at higher 5CB concentrations.

**Introduction**

Initially proposed as a theoretical possibility for bent-core molecules[1–3], the $N_{TB}$ phase of bent mesogenic dimer liquid crystals has recently been observed and intensively studied in several systems of bent molecular dimers[4–14], the most studied of which is CB7CB. In the conventional nematic phase, the ground state exhibits a uniformly aligned director field in space. In contrast, bend deformation of the director field in the $N_{TB}$ phase is energetically favored because of the bent molecular shape of its constituents, with the bent molecular dimers twisting along a helical axis to form a heliconical director field with the same sign and magnitude of bend and twist director deformation everywhere[2]. The usual nematic symmetry is subsequently broken, and the phase breaks spontaneous polar and reflection symmetry, even though the molecules are achiral.

Several noteworthy features are associated with the underlying molecular organization. In a polarized light microscope, the $N_{TB}$ phase usually exhibits focal conic textures that are reminiscent of smectic phases in addition to one-dimensional periodic structures in an untreated glass cell and oriented rope-like textures in unidirectionally rubbed planar cells[8,14,15]. However, no x-ray reflection peaks indicative of smectic order are observed in the $N_{TB}$ phase, suggesting that there is no electron density modulation associated with the periodic structure[4]. The chirality of the $N_{TB}$ phase has been confirmed using NMR spectroscopy[16] and by the observation of an electroclinic effect, with the field response characteristic of a helical structure of nanometer scale pitch[17]. The nanoscale structure of the $N_{TB}$ phase of CB7CB was observed directly using freeze-fracture transmission electron microscopy (FFTEM)[5,7], which revealed periodically modulated structures with a characteristic length of ~8 nm. This modulation, which has a period comparable to the molecular scale, has the shortest observed supermolecular periodicity of any nematic fluid. Similarly nanometer-scale pitches have now been confirmed in several other liquid crystal systems that have the $N_{TB}$ phase[6,18].

The present study of mixtures of the $N_{TB}$-forming bent molecular dimer CB7CB with the rod-like 5CB liquid crystal was motivated by a desire to understand the nature of the $N_{TB}$ helical twist. In this manuscript, we present a systematic study of mixtures of CB7CB and 5CB using polarized light microscopy (PLM), differential scanning calorimetry (DSC), and FFTEM.

**Materials and Methods**

CB7CB (4',4'-(heptane-1,7-diyl)bis(([1',1"-biphenyl]-4"-carbonitrile))) was synthesized as described previously[5] and via another method (Supplementary Information). We purchased 5CB (4-cyano-4′-pentylbiphenyl) from Sigma-Aldrich and used it as received. We prepared mixtures of CB7CB and 5CB by several steps of mechanically mixing the two compounds at 120°C (where they are both isotropic liquids), followed by centrifugation for ~ 1 minute.

The liquid crystal mixtures were filled into Instec 3.2 µm unidirectionally rubbed commercial cells and home-made untreated cells by capillary action at high temperature in the isotropic phase. The samples were typically cooled slowly from the isotropic phase (at ~2°C min$^{-1}$) in order to avoid any hysteresis effects. Temperature was controlled using an Instec STC200D temperature controller. Textural observations were made using a Nikon Eclipse E400 polarizing microscope equipped with an Olympus Camedia C-5050 Zoom digital camera.

We prepared the mixtures for FFTEM experiments by sandwiching them between 2 mm by 4 mm untreated glass slides spaced with a several-micron-thick gap and observed the cell on a hot stage under a polarized light microscope. We heat the cells to ~120°C and then slowly cool into the desired phase. Once the sample has equilibrated at the desired temperature, we eject the cell from the hot stage into a well of liquid propane which rapidly quenches the cell to $T < 90$ K. We then transfer the cell into liquid nitrogen, then finally into a Balzers BAF-060 freeze-fracture apparatus under high vacuum with $T \sim 140$ K, where it is mechanically fractured by pulling the glass plates of the cell apart, exposing a rough nanotextured surface. We shadow the fracture face by obliquely evaporating 2 nm of platinum to capture the topographic structure of the exposed surface. A final coating of 25 nm carbon normal to the surface completes the electron absorption replica, with which the interfacial topography can be viewed and imaged in a transmission electron microscope.

Differential scanning calorimetry measurements were carried out using a Mettler Toledo DSC823e/700.

**Results and Discussion**

We mixed 5CB into CB7CB in different ratios from 12.5% to 95% by weight (the mixtures are denoted $x$% 5CB/CB7CB, where $x$ is the weight percentage of 5CB). We constructed the phase diagram of the 5CB/CB7CB mixtures in Fig. 1 by PLM observations and DSC measurements. The isotropic (Iso) to nematic (N) transition temperature decreases monotonically as the 5CB concentration increases, which indicates nearly ideal mixing in these phases. Interestingly, the N–$N_{TB}$ phase transition temperature decreases almost exactly linearly with the addition of 5CB, up to the last measureable concentration of 62.5%. Although this behavior resembles classical freezing-point depression, the observed decrease in the transition temperature of the mixtures with the addition of 5CB is much smaller in magnitude than expected for conventional freezing point depression.

*Polarized light microscopy study.* We investigated the mixtures using PLM. On transition from the Iso to the N phase, we identify the characteristic Schlieren textures in untreated cells, while rubbed cells show alignment of the liquid crystal and uniform birefringence (Figs. 2a and b). The $N_{TB}$ phase exhibits a broken fan texture of somewhat lower birefringence in untreated glass cells, giving it its smectic-like appearance. In unidirectionally rubbed cells, we observe well-aligned focal conic textures just below the N–$N_{TB}$ transition (Figs. 2a and b). This focal conic texture gives way to a characteristic rope-like or stripe texture which is due to contraction of the $N_{TB}$ helix pitch on decreasing temperature, inducing an undulation instability in the cell[19]. Magnetic field measurements show that this stripe textured arrangement is not the ground state for the $N_{TB}$ system, but that an undulation-free director configuration is[20].

For neat CB7CB and for 5CB/CB7CB mixtures with lower 5CB concentrations (12.5% to 50%) at the N–$N_{TB}$ transition, domains of $N_{TB}$ nucleate from the N phase and quickly grow and coarsen until the phase transition is complete. Figs. 2a and b show N/$N_{TB}$ phase coexistence in neat CB7CB and 25% 5CB/CB7CB samples, respectively. While nematic director fluctuations are clearly visible in the N phase, in the $N_{TB}$ phase the director fluctuations freeze out on cooling and cannot be detected by PLM. For mixtures of higher 5CB concentration, the nucleation of $N_{TB}$ domains occurs rapidly, with the nematic director fluctuations disappearing abruptly at the transition temperature and smoother, well-aligned $N_{TB}$ textures forming, as shown in Fig. 2c. When the 5CB concentration is further increased to 75%, we observe the nematic director fluctuations slowly freeze out down to -20°C, with no other observable indications of a phase transition to the $N_{TB}$ phase (Fig. 2d). This mixture represents the concentration at which twist-bend nematic-dominated phase behavior gives way to 5CB-dominated behavior.

On cooling the mixtures of CB7CB with 5CB in a microscope hot stage, we inevitably create a (mostly) vertical temperature gradient in the cell. On reducing the temperature a few degrees below the N–$N_{TB}$ transition, we observe a bright birefringent stripe texture which aligns along the rubbing direction and fills a large fraction of the cell (Fig. 3a and c). These stripes flash in and out as the temperature settles, and after ~ 5 min they relax away nearly completely (Fig. 3b), such a smooth uniform texture is left. The temperature gradient appears to induce these stripes, as they do not occur at a given temperature. On further cooling into the $N_{TB}$ phase, we observe the stripes fill the cell once again (Fig 3c), but only a small fraction of them relax back into the uniform state at this temperature (Fig. 3d). In the case of 37.5% 5CB/CB7CB, the stripes persist in the phase at ~5°C below the N–$N_{TB}$ transition. The $N_{TB}$ phase becomes more viscous on decreasing temperature[21] and eventually cannot anneal the induced undulations. On increasing 5CB concentration, however, the $N_{TB}$ phase becomes more fluid. We find that in the 62.5% 5CB/CB7CB mixture, at temperatures lower than 12°C below the N–$N_{TB}$ transition, we observe relatively few stripes which relax into the uniform state in a matter of several seconds.

*Calorimetric study.* The latent heat released at the Iso–N and N–$N_{TB}$ transitions are plotted as a function of 5CB weight percentage in Fig. 4. The latent heat of the Iso–N transition increases with increasing 5CB concentration in the mixtures (Fig. 4a). The linearity of the Iso–N transition latent heat peaks in the mixtures with increasing 5CB concentration indicates that the CB7CB and 5CB are nearly ideally miscible in the N phase. We note that the latent heat of the Iso–N transition in CB7CB is about half that of 5CB. The reason for this is that the two "arms" of CB7CB are linked through an alkyl chain which reduces the conformational degrees of freedom available to the molecule, which amounts to a halving of the latent heat of CB7CB with respect to that of 5CB at the Iso–N transition.

The latent heat released in the N–$N_{TB}$ transition, on the other hand, decreases with increasing 5CB concentration (Fig. 4b). This decrease is faster than expected from freezing point depression, based on the relative proportions of CB7CB and 5CB present in the mixtures. When the concentration of 5CB approaches 37.5%, the latent heat of the N–$N_{TB}$ transition is barely detectable in the DSC plots (see Supplementary Fig. S1). This observation is consistent with a calorimetric study on 5CB/CB9CB mixtures[9] which confirms the first-order nature of the neat CB9CB, shows a steady decrease in the latent heat of the N–$N_{TB}$ transition with increasing 5CB concentration, and no evidence from the DSC of the transition above 40% 5CB. The decrease in specific heat associated with the N–$N_{TB}$ transition of the 5CB/CB7CB mixtures along with the PLM observations indicate that the first-order nature of the transition weakens with increasing 5CB concentration,

*Freeze-fracture transmission electron microscopy study.* To probe the nanoscale structure of the $N_{TB}$ phase in the 5CB/CB7CB mixtures, we recorded dozens of FFTEM images of each mixture in the $N_{TB}$ phase (see Supplementary table 1). Typical FFTEM images of CB7CB and several of the 5CB/CB7CB mixtures are shown in Fig. 5. Periodic modulations similar to those found in the $N_{TB}$ phase of neat CB7CB[5,7] are present in mixtures with 5CB concentrations ranging from 12.5% to 62.5%. In a 75% 5CB/CB7CB mixture quenched at -20°C, we observed no clear $N_{TB}$-like periodic modulations. In both neat CB7CB and in 5CB/CB7CB mixtures, neither stripe alternation nor half-order reflections in the Fourier transforms of the modulation patterns are present in the FFTEM images, which suggest that the observed periodicity *p* corresponds to the pitch of the twist-bend helix in the mixtures. Based on the similarity of these topographical features to the modulations observed in neat CB7CB, we conclude that the modulations observed in the mixtures represent the intersection of the 3D heliconical structure of the $N_{TB}$ phase with the fracture surface. However, we find a variety of different pitch lengths within each $N_{TB}$ domain in the FFTEM images. This variation of the observed period *p* indicates that we must consider several experimental effects, as follows. When the fracture plane **f** is perpendicular to the viewing direction **v** (the incident electron beam propagation direction) and the $N_{TB}$ helix **n** lies on the fracture plane, value of $p$ corresponds precisely to the ground state pitch $p_0$ (Fig. 6a). When $\mathbf{f} \perp \mathbf{v}$ and the $N_{TB}$ helix **n** does not lie in the fracture plane **f**, the observed period $p$ is related to the true period $p_0$ by $p(\psi) = p_0/\cos\psi$, where $\psi$ is the angle between **n** and **f** (Fig. 6b). On the other hand, when **f** is *not* perpendicular to **v**, we may observe values of the pitch which are smaller than $p_0$ (Fig. 6c). In practice, the fracture plane **f** is almost always nearly perpendicular to the view direction **v**, but this effect contributes some anomalously small values to the observed pitch distribution (Supplementary Fig. S2), which have also been observed in another FFTEM study of the $N_{TB}$ phase[7]. In addition to these effects we expect the sample, and therefore the $N_{TB}$ domains, to thermally contract by a small percent during quenching (~1 to 5%), tending to decrease the measured value of the pitch with respect to that before quenching. With these considerations in mind, we implemented a statistical approach to measuring the FFTEM images of the $N_{TB}$ pitch in the mixtures.

We measure the pitch within a given $N_{TB}$ domain by taking a spatial Fourier transform of the region (Fig. 5), and recording its area. We use the area as a weighting factor for the corresponding measured pitch value. This process is carried out for each $N_{TB}$ domain in each image, and the data are plotted in a weighted histogram. In principle, the majority of the fracture area which we observe will exhibit a pitch value which corresponds very nearly to $p_0$, with a distribution around this value which is due to the effects described above, as depicted in Fig. 6 and explained further in Supplementary Fig. S3.

Using this methodology, we measured the temperature dependence of the pitch of neat CB7CB (Fig. 6d). When we quench CB7CB at 100°C, just below the N–$N_{TB}$ transition, we measure a broad pitch distribution which exhibits prominent peaks at ~8 nm and ~9 nm, but with a tail that extends out to ~13 nm. On further cooling, this tail becomes smaller, until only a single peak centered around 8 nm remains at room temperature.

We note that on quenching CB7CB from a variety of different temperatures, the textures within the cell differ from the textures before the quench (Supplementary Fig. S4), but they are not crystallized (Supplementary Fig. S5). The images taken after quenching indicate that thermal contraction may shrink the measured pitch values by a small percentage. Even when quenching from the isotropic phase of CB7CB, FFTEM images and PLM show domains of twist-bend modulations of ~8 nm coexisting with amorphous domains (Supplementary Fig. S6). This indicates that we are quenching into the $N_{TB}$ phase from the isotropic phase, and timescale for formation of the phase from the isotropic is <10 ms, a very quick process (Supplementary Fig. S7).

We investigated the temperature dependence of a 25% 5CB/CB7CB mixture to discern any influence that the 5CB may have on the $N_{TB}$ pitch as a function of temperature (Fig. 6e). Just below the N–$N_{TB}$ phase transition, we measure a very broad pitch distribution, with values ranging from ~7 – 14 nm. This behavior is similar to that observed in the neat CB7CB. Several degrees below the N–$N_{TB}$ transition, the pitch distribution tightens around $p$ = ~8.5 nm. This behavior persists until 29°C, where the pitch distribution broadens once again such that there are several peaks over a broad distribution. We attribute this behavior to the 5CB in the mixture, since it does not occur in the neat CB7CB (Fig. 6d). We do not observe phase separation of the mixture components in the cell at any temperature, which is evidence that the 5CB disrupts the normal helix morphology and induces a change which manifests as large fluctuations in the measured pitch.

We compare the pitch distributions of a series of mixtures of 5CB and CB7CB to investigate the concentration dependence of the ground state pitch. The samples were quenched ~10°C below the N–$N_{TB}$ phase transition in the respective mixtures to account for any possible temperature dependence of the pitch. Intriguingly, we found that the pitch depends on the 5CB concentration. In mixtures up to 37.5% 5CB/CB7CB, the pitch distribution shifts modestly toward larger pitch values on increasing 5CB concentration, from 7.99 ± 0.14 nm in neat CB7CB to 9.20 ± 0.30 nm in the 37.5% 5CB/CB7CB mixture, as shown in Fig. 6f. We conclude that the observed changes in the pitch distributions across these mixtures are principally a result of swelling of the $N_{TB}$ helix by 5CB.

At 50% 5CB/CB7CB concentration, the pitch distribution broadens, with prominent peaks at several different pitch values (Fig. 6e). At 62.5% 5CB/CB7CB concentration, we find a broad peak which is centered at around ~18 nm. This behavior implies the presence of a discontinuous structure transition in the pitch around the 50% 5CB/CB7CB concentration. This phenomenon may indicate a discontinuous change in the twist-bend helix or even a transition to a unique liquid crystal structure with an 18 nm modulation. Further study will be necessary to determine its exact nature.

We observe nanometer-scale modulations in all the mixtures until we reach concentrations of 75% 5CB/CB7CB and above. Even when the 75% 5CB/CB7CB mixture is quenched at -20°C, a temperature which should roughly correspond to the $N_{TB}$ phase according to an extrapolation from Fig. 1, we do not observe clear nanometer-scale modulations in FFTEM.

On varying the tail length $n = 5 - 8$ of the nematic dopant in 50% $n$CB/CB7CB mixtures, we find only a very subtle apparent decrease in the position of the pitch distribution, as shown in Supplementary Fig. S8.

**Conclusions**

The achiral, odd-methylene-linked dimer CB7CB and the rod-like liquid crystal 5CB form nearly ideal mixtures in the Iso and N phases, and are well-mixed in the $N_{TB}$ phase, at concentrations as high as 62.5% 5CB/CB7CB. FFTEM reveals nanometer-scale modulations in these mixtures which are nearly identical to those which are observed in neat CB7CB. We utilized a statistical approach to analyze the FFTEM images and study the temperature dependence of the pitch in neat CB7CB and its mixtures with 5CB and the concentration dependence of the pitch in the mixtures. We identify a structural transition that occurs at 50% 5CB/CB7CB concentration which remains a mystery.

**Acknowledgement**

This work was supported by the Soft Materials Research Center under NSF MRSEC Grants DMR-0820579 and DMR-1420736, by Institute for Complex Adaptive Matter Postdoctoral Fellowship Award OCG5711B, and by ED GAANN Award P200A120014.

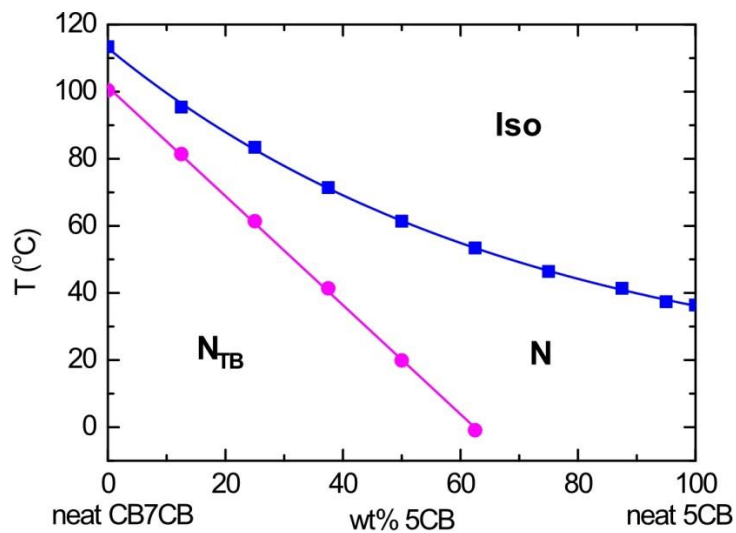

**Figure 1. Phase diagram of mixtures of CB7CB and 5CB.** Both the Iso–N and N–$N_{TB}$ transition temperatures decrease with the addition of 5CB, with the N–$N_{TB}$ phase transition temperature decreasing practically linearly. These values are obtained by DSC except for the temperatures measured for the N–$N_{TB}$ transition at 37.5%, 50%, and 62.5% 5CB concentrations, which were measured by PLM.

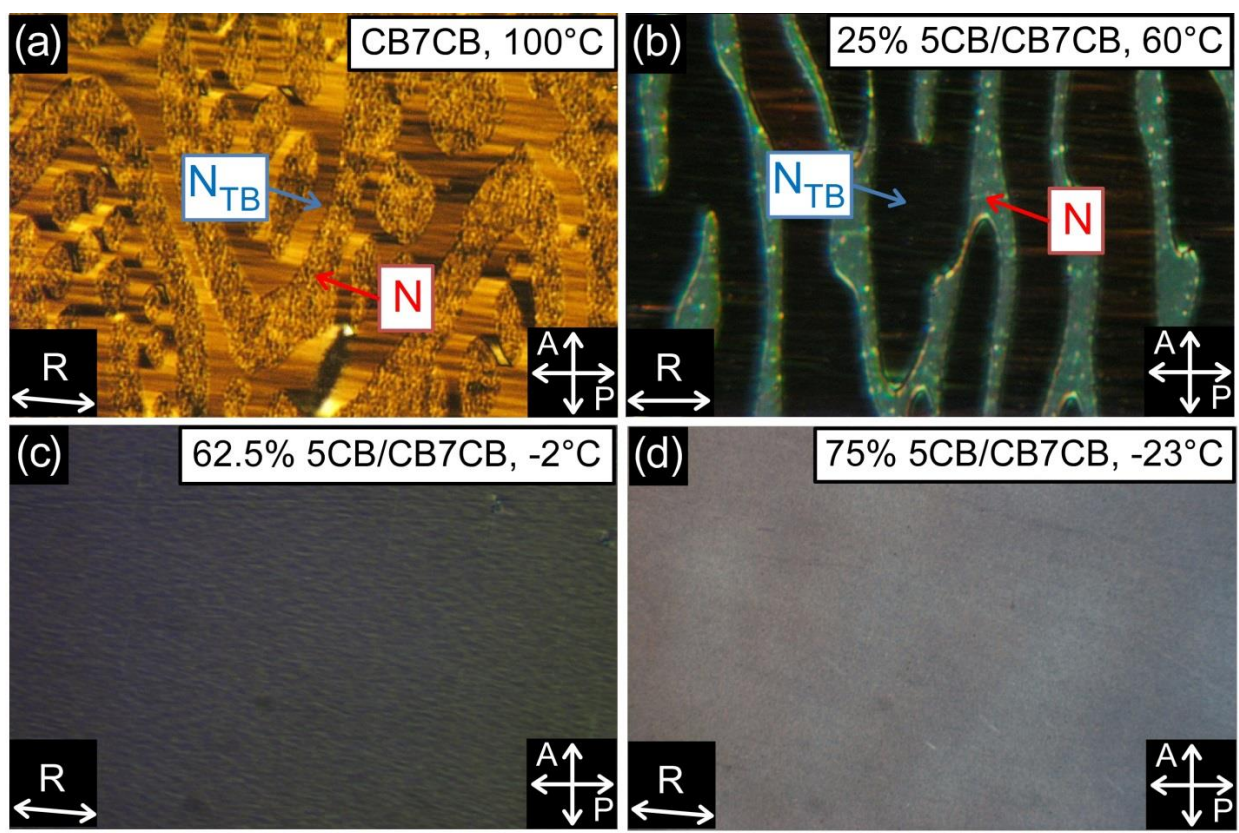

**Figure 2. Optical textures of neat CB7CB and 5CB/CB7CB mixtures near the N–$N_{TB}$ transition in unidirectionally aligned planar cells.** In neat CB7CB (a) and 25% 5CB/CB7CB (b), we see N–$N_{TB}$ phase coexistence when the samples are cooled slowly from the nematic, confirming the first order natures of the N–$N_{TB}$ phase transition. (c) In a 62.5% 5CB/CB7CB mixture, a weak stripe texture appears when we cool 2°C below the N–$N_{TB}$ transition. In this mixture we find no evidence of phase coexistence, indicating that the first-order nature of the transition weakens with increasing 5CB concentration. (d) As a 75% 5CB/CB7CB cell is cooled to a temperature below the putative transition temperature as extrapolated from the phase diagram in Fig. 1 (~-22°C), the nematic director fluctuations weaken and eventually become unobservable. With the available evidence we cannot determine the nature of this phase.

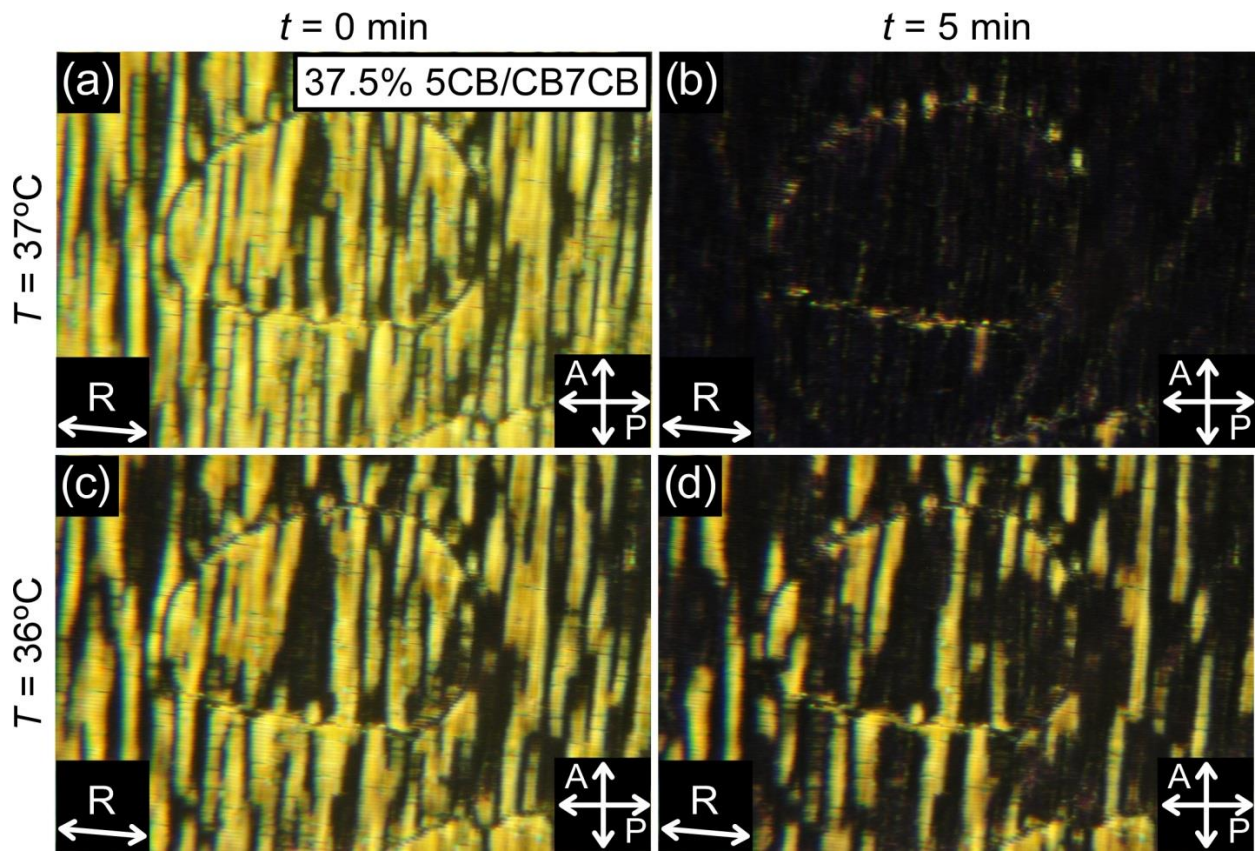

**Figure 3. Annealing of stripe textures in the 37.5% 5CB/CB7CB mixture.** On cooling at 1°C min$^{-1}$ from 38°C to 37°C, a transient stripe texture appears (a), which anneals into a (mostly) uniformly aligned state after 5 minutes (b). On further cooling by 1°C min$^{-1}$ from 37°C to 36°C, stripes appear once again (c). Because of the increase in the viscosity of the mixture on cooling, the texture does not completely anneal into the uniform ground state (d). Further cooling produces more stripes which do not anneal away, until the cell is completely filled with them.

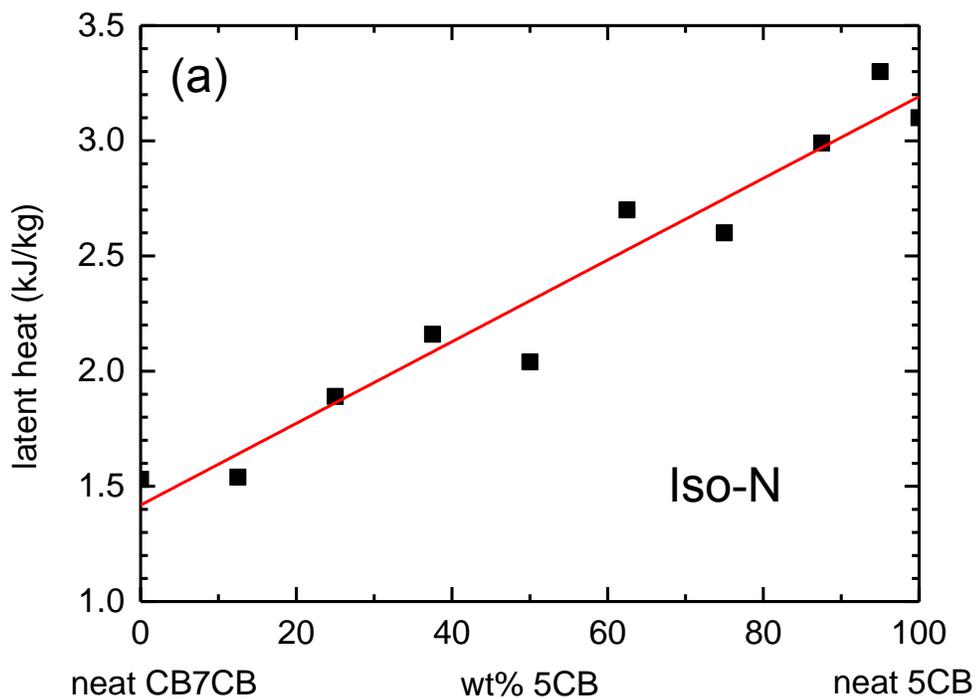

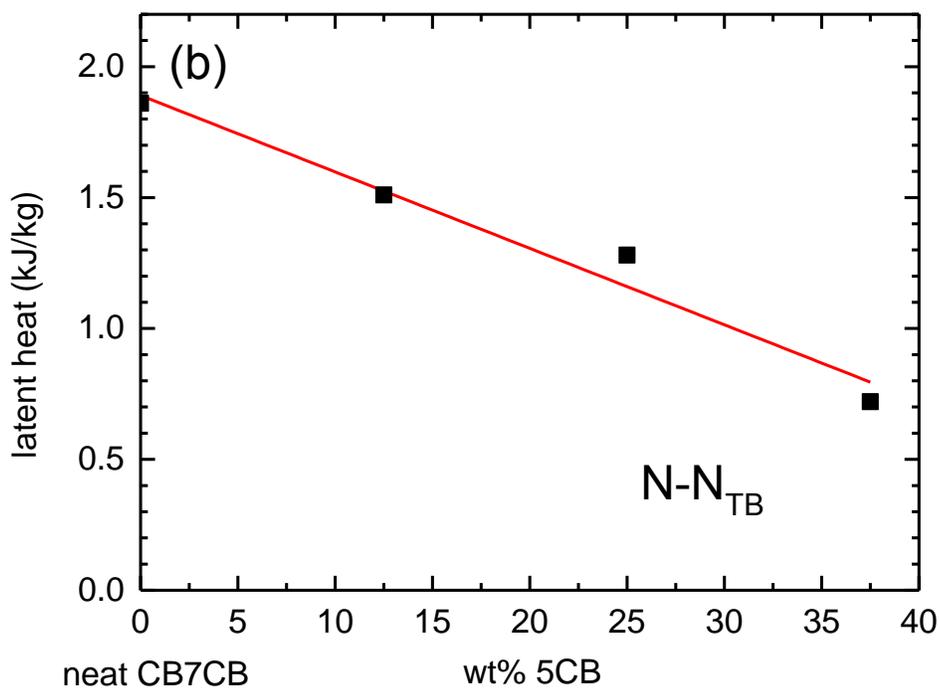

**Figure 4. Latent heat of the Iso–N and N–N$_{TB}$ transitions in mixtures of CB7CB and 5CB.** (a) The latent heat of the Iso–N transition increases continuously with 5CB concentration, reflecting the larger latent heat content of the Iso–N transition of neat 5CB. (b) The latent heat of the N–N$_{TB}$ transition in the mixtures decreases more quickly than expected based on the concentration of the components alone. By extrapolating a linear fit, a tricritical point of the mixtures should exist at ~60% 5CB concentration.

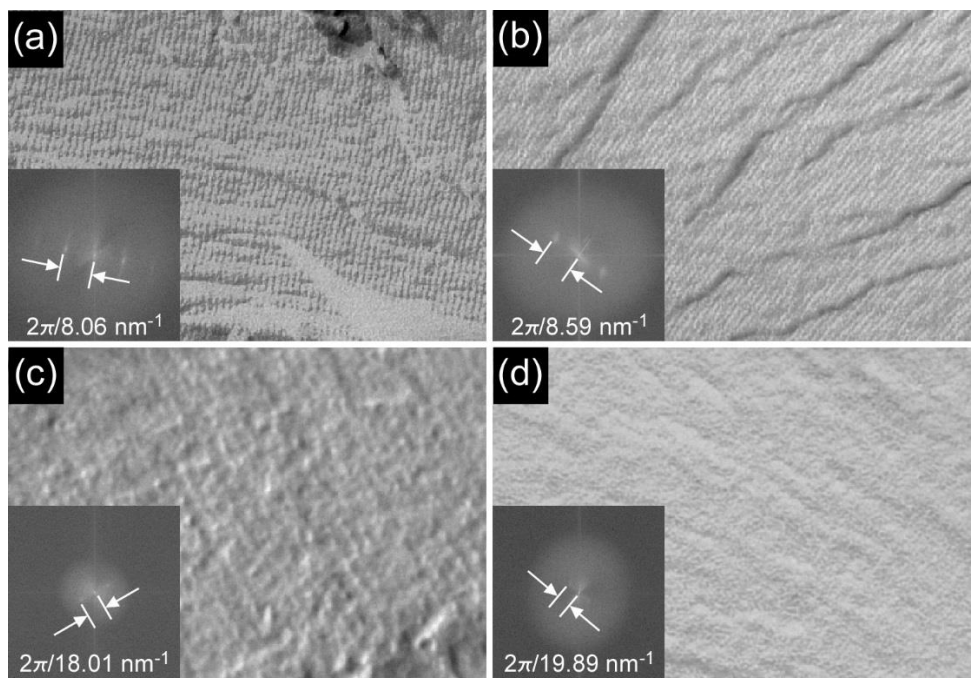

**Figure 5. FFTEM images showing twist-bend modulations in (a) neat CB7CB, (b) 25% 5CB/CB7CB, (c) 50% 5CB/CB7CB, and (d) 62.5% 5CB/CB7CB.** The pitch is measured using spatial Fourier transforms, which are shown as insets. As the 5CB concentration increases in the mixtures, the twist-bend modulations are generally of larger spacing and become visibly weaker. All the images are at the same scale.

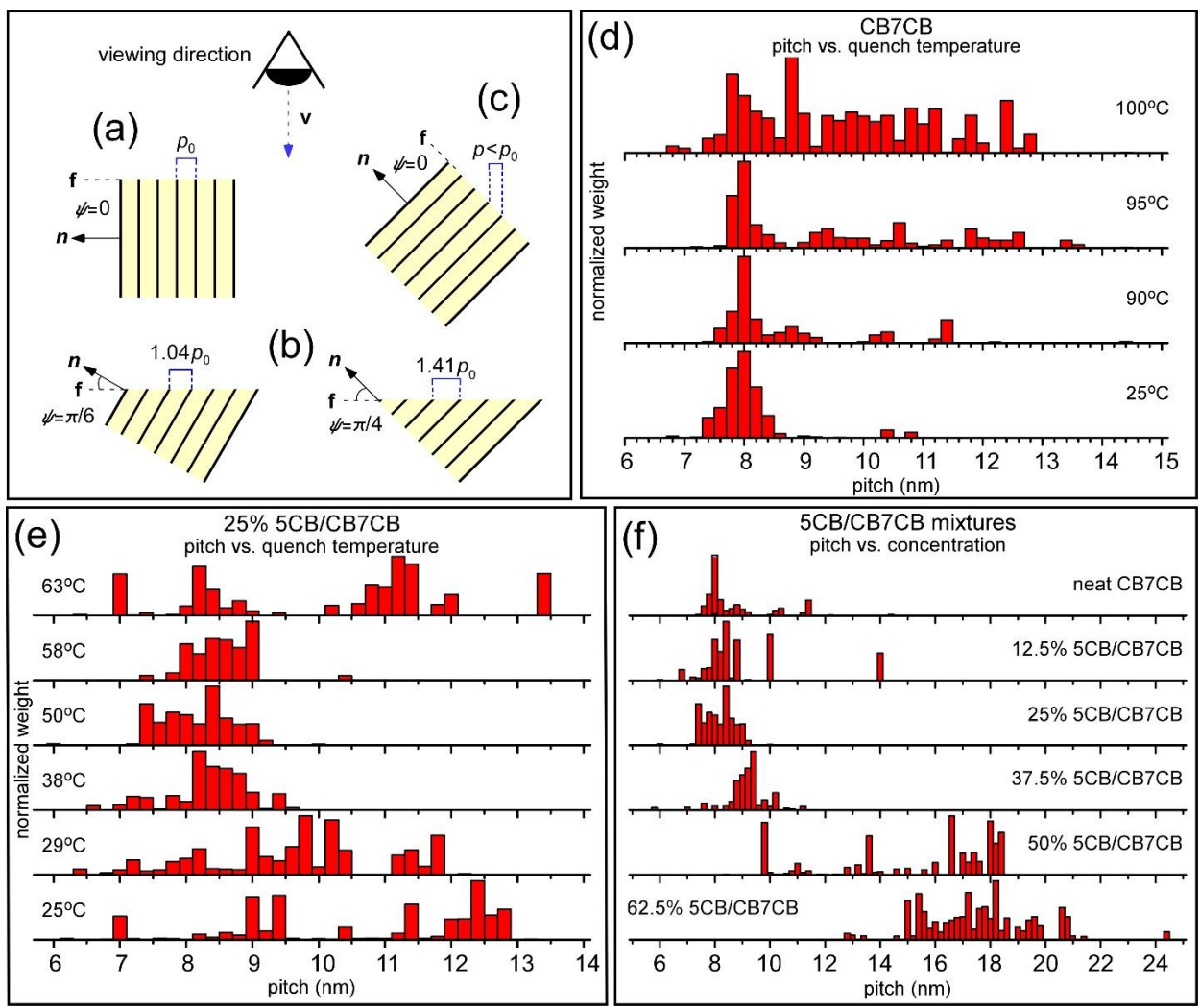

**Figure 6. Schematic and pitch measurements of the $N_{TB}$ phase of CB7CB and 5CB/CB7CB mixtures from FFTEM.** (a-c) Schematics of the possible configurations of twist-bend 'pseudo-layered' domains in FFTEM with helix axis **n** as viewed from viewing direction **v** onto fracture plane **f**, and assuming that the pseudo-layer spacing is fixed at the value $p_0$. (a) When **n** in the plane **f** and **f** $\perp$ **v**, we observe the ground state pitch $p_0$ of the $N_{TB}$ phase. (b) When **n** makes an arbitrary angle $\psi$ with respect to **f** and **f** $\perp$ **v**, we observe a pitch $p > p_0$ which is described by the equation $p(\psi) = p_0/\cos\psi$. (c) When **n** // **f** and the **f** makes an arbitrary angle with respect to **v**, we observe a pitch $p < p_0$. (d-f) The weighted histograms are obtained by measuring the pitch in each domains in the FFTEM images of a given preparation and weighting them by the area of the domain. The bin size is 0.2 nm in all histograms. (d) Neat CB7CB pitch distributions as a function of quenching temperature. At high quenching temperature, the pitch distribution exhibits

prominent peaks at ~8 nm and ~9 nm and has a very broad tail which extends out to ~13 nm. As we reduce the quenching temperature, the tail becomes smaller until it essentially vanishes at room temperature. (e) The pitch distribution of the 25% 5CB/CB7CB mixture as a function of quenching temperature. At quenching temperatures just below the N–N$_{TB}$ phase transition, we find a broad pitch distribution, a behavior similar to that of neat CB7CB. On further reducing the quench temperature, the pitch distribution tightens and does not vary much with temperature. At quenching temperature 29°C, the pitch distribution broadens once again. In contrast to the behavior of neat CB7CB (d), we find that the 5CB influences the pitch in the mixture at low quenching temperature, such that the pitch distribution broadens and exhibits multiple peaks. (f) Pitch distributions of several 5CB/CB7CB mixtures. The pitch increases steadily from ~8 nm in neat CB7CB to ~9.2 nm in the 37.5% 5CB/CB7CB mixture. At 50% 5CB/CB7CB, however, the pitch becomes unexpectedly broad, and spans from ~9 nm to ~18 nm. The 62.5% 5CB/CB7CB mixture exhibits a Gaussian-like distribution centered at ~18 nm, albeit much broader than those of the mixtures with smaller concentrations of 5CB. This behavior indicates a structural transition occurring at around 50% 5CB/CB7CB. Further study will be necessary to determine the nature of the transition.

**References**

1. Meyer, R. Structural problems in liquid crystal physics. *Houches Summer Sch. Theor. Phys.* 273–373 (1973).

2. Dozov, I. On the spontaneous symmetry breaking in the mesophases of achiral banana-shaped molecules. *EPL Europhys. Lett.* **56,** 247 (2001).

3. Memmer, R. Liquid crystal phases of achiral banana-shaped molecules: a computer simulation study. *Liq. Cryst.* **29,** 483–496 (2002).

4. Cestari, M. *et al.* Phase behavior and properties of the liquid-crystal dimer 1″,7″-bis(4-cyanobiphenyl-4′-yl) heptane: A twist-bend nematic liquid crystal. *Phys. Rev. E* **84,** (2011).

5. Chen, D. *et al.* Chiral heliconical ground state of nanoscale pitch in a nematic liquid crystal of achiral molecular dimers. *Proc. Natl. Acad. Sci.* **110,** 15931–15936 (2013).

6. Wang, Y. *et al.* Room temperature heliconical twist-bend nematic liquid crystal. *CrystEngComm* **17,** 2778–2782 (2015).

7. Borshch, V. *et al.* Nematic twist-bend phase with nanoscale modulation of molecular orientation. *Nat. Commun.* **4,** (2013).

8. Tamba, M. G., Baumeister, U., Pelzl, G. & Weissflog, W. Banana-calamitic dimers: unexpected mesophase behaviour by variation of the direction of ester linking groups in the bent-core unit. *Liq. Cryst.* **37,** 853–874 (2010).

9. Tripathi, C. S. P. *et al.* Nematic-nematic phase transition in the liquid crystal dimer CBC9CB and its mixtures with 5CB: A high-resolution adiabatic scanning calorimetric study. *Phys. Rev. E* **84,** (2011).

10. Mandle, R. J., Voll, C. C. A., Lewis, D. J. & Goodby, J. W. Etheric bimesogens and the twist-bend nematic phase. *Liq. Cryst.* **0,** 1–9 (2015).

11. Gorecka, E. *et al.* A Twist-Bend Nematic (NTB) Phase of Chiral Materials. *Angew. Chem.* **127,** 10293–10297 (2015).

12. Jansze, S. M., Martínez-Felipe, A., Storey, J. M. D., Marcelis, A. T. M. & Imrie, C. T. A Twist-Bend Nematic Phase Driven by Hydrogen Bonding. *Angew. Chem. Int. Ed.* **54,** 643–646 (2015).

13. Adlem, K. *et al.* Chemically induced twist-bend nematic liquid crystals, liquid crystal dimers, and negative elastic constants. *Phys. Rev. E* **88,** (2013).

14. Henderson, P. A. & Imrie, C. T. Methylene-linked liquid crystal dimers and the twist-bend nematic phase. *Liq. Cryst.* **38,** 1407–1414 (2011).

15. Panov, V. P. *et al.* Spontaneous Periodic Deformations in Nonchiral Planar-Aligned Bimesogens with a Nematic-Nematic Transition and a Negative Elastic Constant. *Phys. Rev. Lett.* **105,** (2010).

16. Beguin, L. *et al.* The Chirality of a Twist–Bend Nematic Phase Identified by NMR Spectroscopy. *J. Phys. Chem. B* **116,** 7940–7951 (2012).

17. Meyer, C., Luckhurst, G. R. & Dozov, I. Flexoelectrically Driven Electroclinic Effect in the Twist-Bend Nematic Phase of Achiral Molecules with Bent Shapes. *Phys. Rev. Lett.* **111,** (2013).

18. Chen, D. *et al.* Twist-bend heliconical chiral nematic liquid crystal phase of an achiral rigid bent-core mesogen. *Phys. Rev. E* **89,** (2014).

19. Clark, N. A. & Meyer, R. B. Strain-induced instability of monodomain smectic A and cholesteric liquid crystals. *Appl. Phys. Lett.* **22,** 493–494 (1973).

20. Challa, P. K. *et al.* Twist-bend nematic liquid crystals in high magnetic fields. *Phys. Rev. E* **89,** (2014).

21. Salili, S. M. *et al.* Flow properties of a twist-bend nematic liquid crystal. *RSC Adv* **4,** 57419–57423 (2014).

22. Costello, M. J., Fetter, R. & Höchli, M. Simple procedures for evaluating the cryofixation of biological samples. *J. Microsc.* **125,** 125–136 (1982).

23. NIST - properties of methane, ethane, propane, etc.

24. Yang, G., Migone, A. D. & Johnson, K. W. Heat capacity and thermal diffusivity of a glass sample. *Phys. Rev. B* **45,** 157 (1992).

25. Marinelli, M., Mercuri, F., Zammit, U. & Scudieri, F. Thermal conductivity and thermal diffusivity of the cyanobiphenyl (n CB) homologous series. *Phys. Rev. E* **58,** 5860 (1998).

26. Mansaré, T., Decressain, R., Gors, C. & Dolganov, V. K. Phase Transformations And Dynamics Of 4-Cyano-4′-Pentylbiphenyl (5cb) By Nuclear Magnetic Resonance, Analysis Differential Scanning Calorimetry, And Wideangle X-Ray Diffraction Analysis. *Mol. Cryst. Liq. Cryst.* **382,** 97–111 (2002).

27. Decressain, R., Cochin, E., Mansare, T. & More, M. Polymorphism and dynamics of MBBA as studied by NMR. *Liq. Cryst.* **25,** 517–523 (1998).

**Supplementary Information**

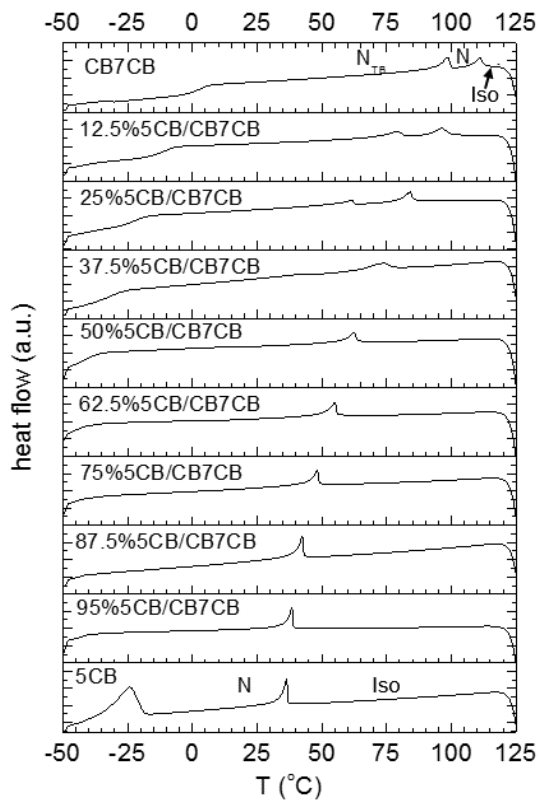

**Supplementary Figure S1. DSC cooling curves of neat CB7CB, neat 5CB, and their mixtures, obtained at a cooling rate of -10°C/min.** In neat CB7CB and in mixtures with less than 37.5% 5CB, we see peaks corresponding to both the Iso–N and N–$N_{TB}$ transitions. Mixtures with higher 5CB concentrations display only an Iso–N transition peak.

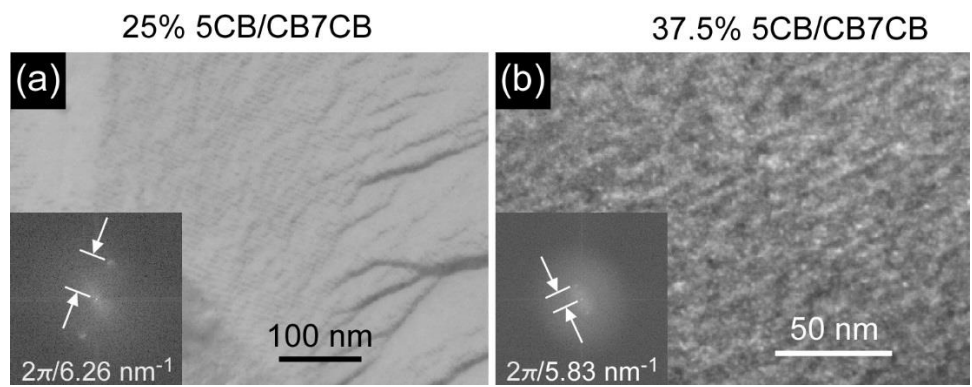

**Supplementary Figure S2. FFTEM images of domains with anomalously small pitch in mixtures of CB7CB and 5CB.** (a and b) When the viewing direction **v** makes an arbitrary angle with respect to the fracture plane **f** (see Fig. 6c), the observed pitch may be less than $p_0$. We tend to observe these smaller pitch values on features of the replica which appear visibly sloped. (a) Domain with a measured pitch of ~6.3 nm in a 25% 5CB/CB7CB mixture. (b) Domain with a measured pitch of ~5.8 nm in a 37.5% 5CB/CB7CB mixture.

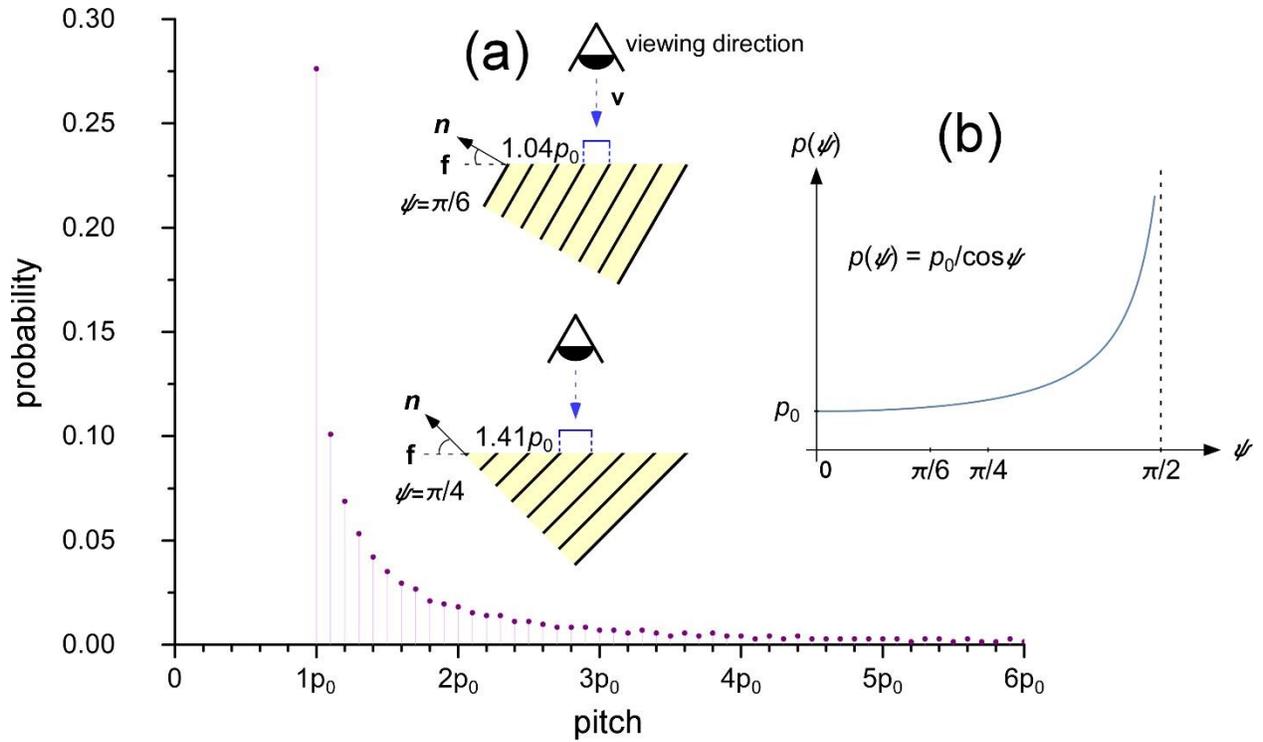

**Supplementary Figure S3. Probability of finding a given twist-bend pitch in FFTEM.** We assume a perfectly random distribution of orientations of $N_{TB}$ domains **n** within a sample, a fracture plane **f** perpendicular to the viewing direction **v**, and a pseudo-layer spacing fixed at $p_0$. (inset (a)). The equation $p = p_0/\cos\psi$ gives the possible values of the observed pitch $p$ as a function of the orientation angle $\psi$ of an $N_{TB}$ domain. As the angle between **n** and **v** becomes smaller, the observed pitch $p$ shifts to larger values (inset (b)). The main plot indicates the probability of observing a given pitch value $p$ in a bin of bounds $p$ and $p + 0.1p_0$. Because the twist-bend helix axis tends to lie in the plane of an untreated glass cell used during FFTEM, the probability distribution is even more strongly peaked at $p_0$.

before quench in liquid propane | after quench in liquid propane

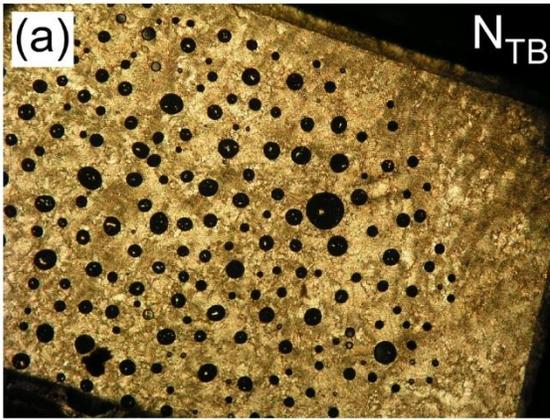 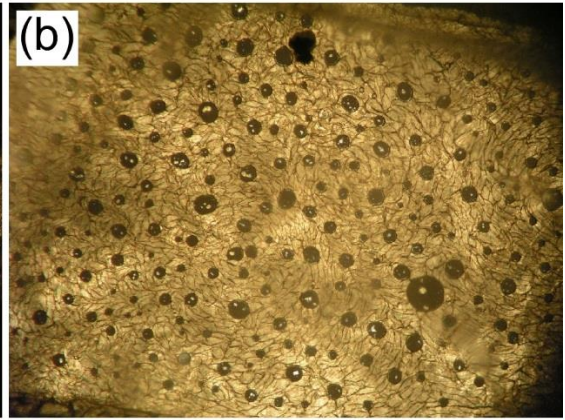

$T_{N-N_{TB}} - T = 76\ °C$ — (a) $N_{TB}$ / (b)

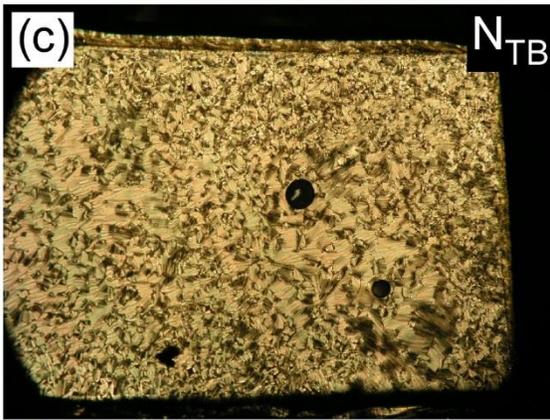 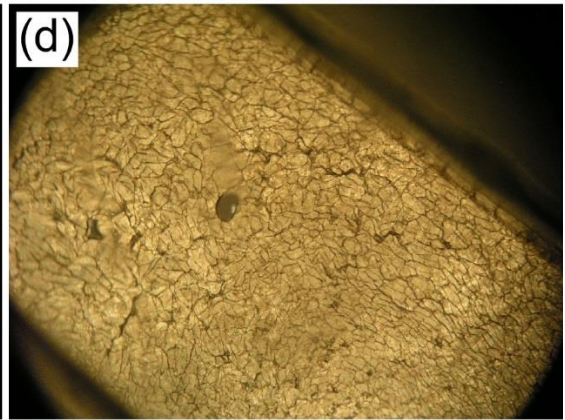

$T_{N-N_{TB}} - T = 0.1\ °C$ — (c) $N_{TB}$ / (d)

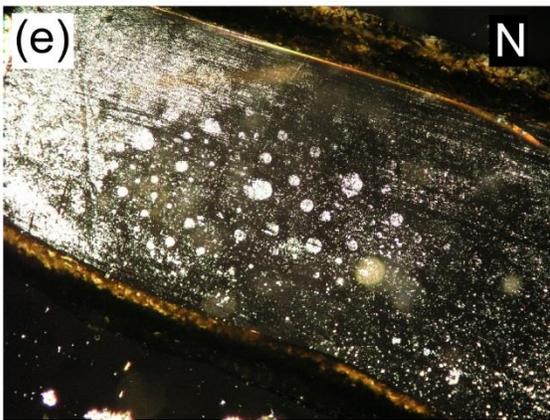 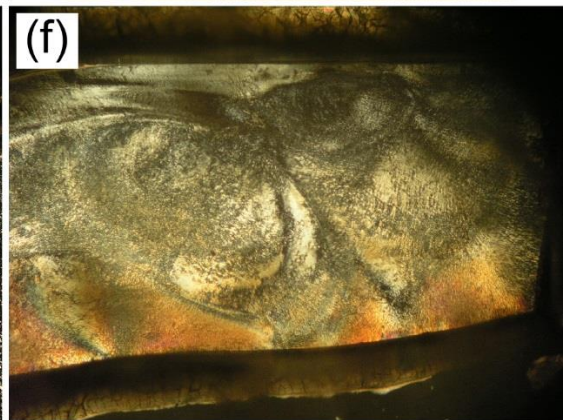

$T_{I-N} - T = 0.1\ °C$ — (e) N / (f)

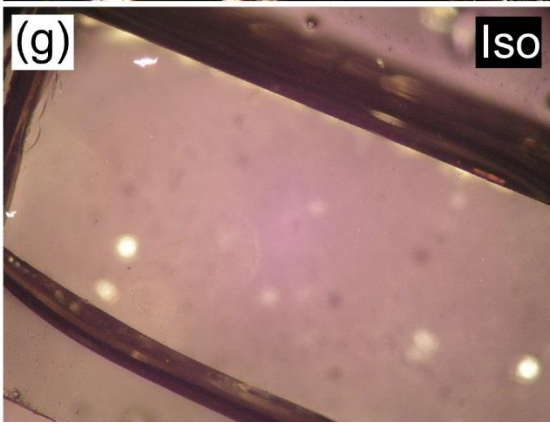 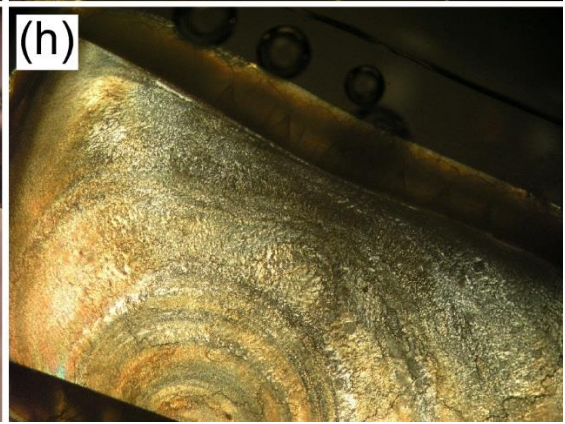

$T - T_{I-N} = 87\ °C$ — (g) Iso / (h)

**Supplementary Figure S4. Polarized optical microscopy images of CB7CB before and after quenching in liquid propane.** We cool glass planchette cells filled with CB7CB to the desired temperature under the microscope, take an image of the texture, then quench the cell in liquid propane (90 K). We then reduce the temperature of the microscope stage to ~130 K, place the cell back under the microscope, and observe the textures. (a) 76°C below the N-$N_{TB}$ transition, the phase exhibits disordered rope-like textures in the untreated glass cell. (b) After rapidly quenching the cell, we recover a texture and birefringence similar to that in (a), except that thermal contraction leaves cracks in the sample. (c) At 0.1°C below the N–$N_{TB}$ transition, we observe focal conic textures of the $N_{TB}$ phase which have not yet been disordered by further cooling. (d) After quenching, we do not recover the original texture but one similar to that observed in (b). (e) 0.1°C below the I–N transition, the nematic is mostly homeotropic in the cell. (f) After quenching, the cell appears to have more birefringence. Based on FFTEM experiments in this temperature range, in which we observed twist-bend modulations of ~8 nm, we believe that a significant fraction of this cell is filled with the $N_{TB}$ phase. (g) 87°C above the clearing point of CB7CB, we observe a featureless isotropic sample. (h) After quenching, we find grainy birefringent textures coexisting with some dark regions. Based on FFTEM experiments in this temperature range in which we observe large, micron-scale domains with 8 nm modulations and a comparable amount of amorphous domains, we believe that these grainy domains are made of the $N_{TB}$ phase, while the dark regions are either isotropic or nematic. This shows that from the isotropic phase, we do not quench quickly enough to suppress the nucleation and growth of micron-scale $N_{TB}$ domains.

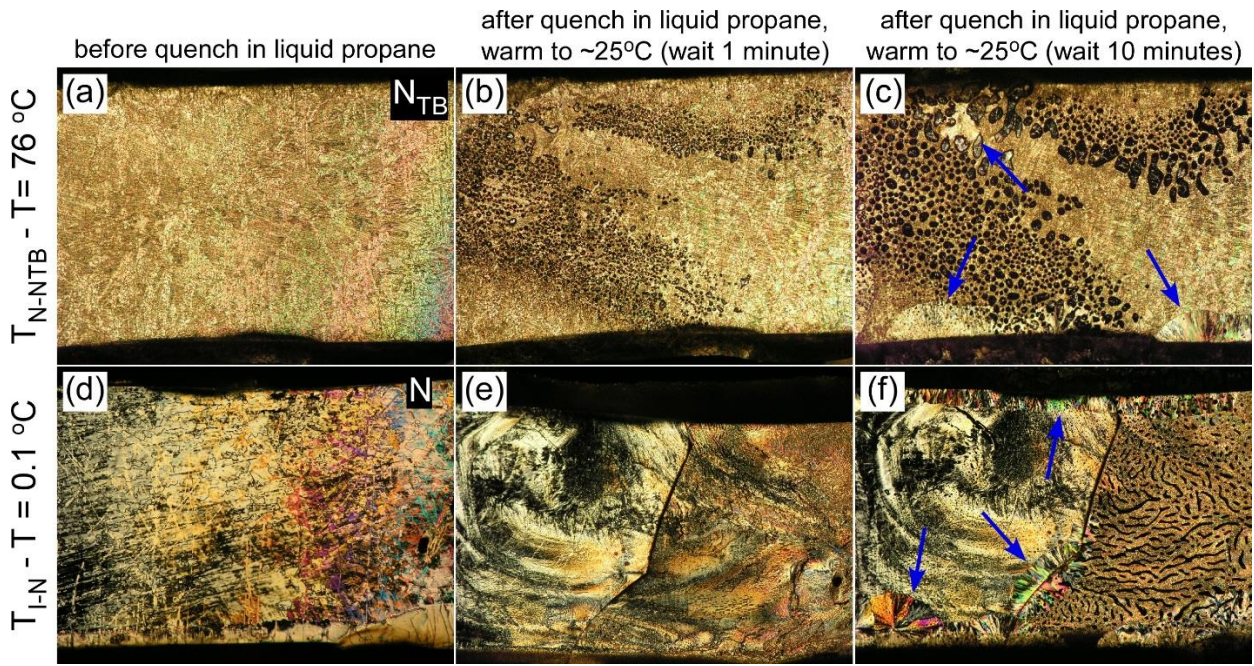

**Supplementary Figure S5. Polarized optical microscopy images of CB7CB before and after quenching and subsequent and warming to room temperature.** We bring glass planchette cells filled with CB7CB to the desired temperature under the microscope, take an image of the texture, then quench the cell in liquid propane (90 K). We then place the cell back into the microscope stage, which is maintained at 25°C, and observe the textures after 1 minute and 10 minutes. (a) 76°C below the N-$N_{TB}$ transition, we find a disordered, rope-like texture in the untreated glass cell. (b) After quenching and subsequent warming to room temperature and waiting 1 minute, we find a similar texture to that in (a) but with air bubbles forming throughout the cell. These come from cracks which form on thermal contraction of the phase during quenching (see Supplementary Fig. S4). (c) After 10 minutes, spherulitic crystal domains begin to grow. (d) 0.1°C below the I-N transition, the nematic exhibits director fluctuations and characteristic Schlieren textures. (e) After quenching and subsequent warming to room temperature and waiting 1 minute, we find that the texture looks very different from (d), and there is a prominent crack in the glass from the rapid quenching. (f) After 10 minutes at room temperature, spherulitic crystal domains begin to grow at the edges of the sample. This shows that the quenching is rapid enough to suppress crystallization during FFTEM.

However, when we go to quenching higher temperatures in CB7CB we do not recover the same textures on warming.

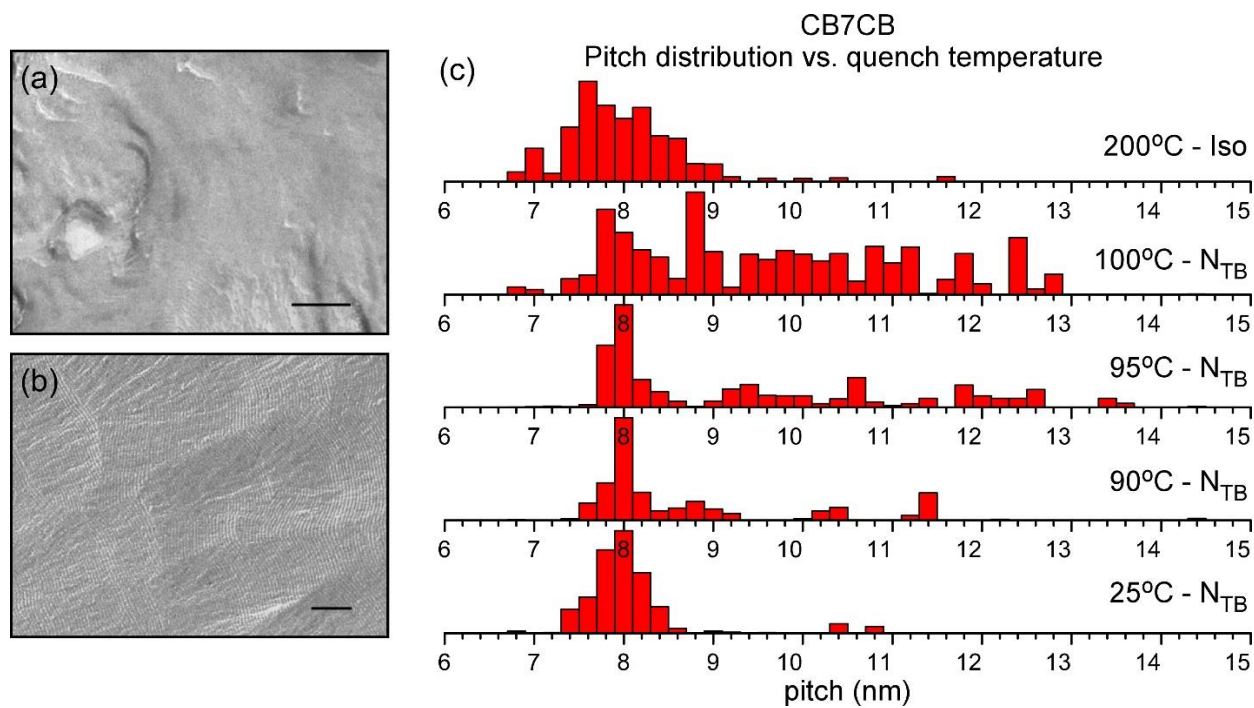

**Supplementary Figure S6. Twist-bend pitch of CB7CB observed by FFTEM from quenching in the isotropic phase of CB7CB.** On quenching CB7CB from 200°C using *copper* planchettes, (87°C above the clearing point), we find some domains which are amorphous (a) and some which contain well-oriented twist-bend modulations of ~8 nm spacing (b). The pitch distribution, measured as described in the main text, peaks around 7.9 nm (c). This demonstrates that the $N_{TB}$ phase forms faster than we can quench the sample with liquid propane. Since the cooling rate for copper planchettes dropped into liquid propane is ~$10^4$ K·s$^{-1}$ (ref. 22), and to vitrify the isotropic phase to a glassy state we need to cool on the order of ~100 K, we place an upper limit on the timescale of formation of micron-scale, well-oriented $N_{TB}$ domains in CB7CB at <10 ms.

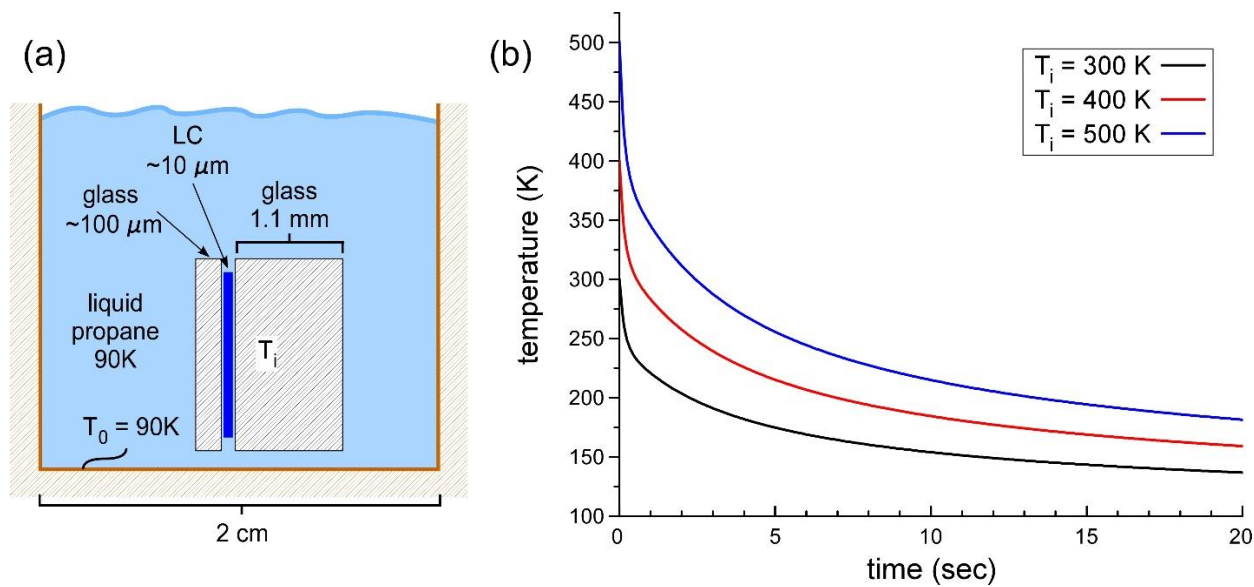

**Supplementary Figure S7. Modeling the transient heat flow in an FFTEM cell.** (a) Schematic of the FFTEM quenching geometry. The cell is immersed in a 2 cm wide reservoir of liquid propane, which is maintained at ~90 K by liquid nitrogen. The cell is composed of two glass substrates of thickness 100 μm and 1.1 mm respectively, with the liquid crystal in between. The entire cell is maintained at the starting temperature until it contacts the liquid propane. We use these experimental values and the 1D heat equation to approximate the transient heat flow in the cell. The thermal diffusivity of liquid propane at ~90 K is $\alpha \sim 10^{-7}$ m² s$^{-1}$ (ref. 23), and that of glass and typical liquid crystals at room temperature is $\alpha \sim 10^{-7}$ m² s$^{-1}$ (refs. 24 and 25). We therefore assume that the whole cell is made entirely of a slab of thermal diffusivity $\alpha = 1 \cdot 10^{-7}$ m² s$^{-1}$, and that κ does not vary with temperature. We believe these to be reasonable approximations given that the thermal diffusivity of glass increases by only ~25% from 300 K to 100 K, and that the liquid crystal sample makes up a small portion of the whole cell. The 1D slab is maintained at 90 K on the boundaries at all times. The initial temperature of the slab is $T = T_i$ at the center (corresponding to the length of the cell), and 90 K everywhere else. We then calculate the temperature at the position of the liquid crystal as a function of time. (b) The temperature as a function of time for three initial temperatures of the cell. The average cooling rate of the cell after the first second of quenching is ~150 K·s$^{-1}$ for $T_i = 500$ K, while it is ~75 K·s$^{-1}$ for $T_i = 300$ K. Liquid crystal 5CB vitrifies at cooling rates equal to or greater than ~0.02 K·s$^{-1}$ (ref. 26), and MBBA vitrifies at cooling rates equal to or greater than ~0.67 K·s$^{-1}$ (ref. 27).

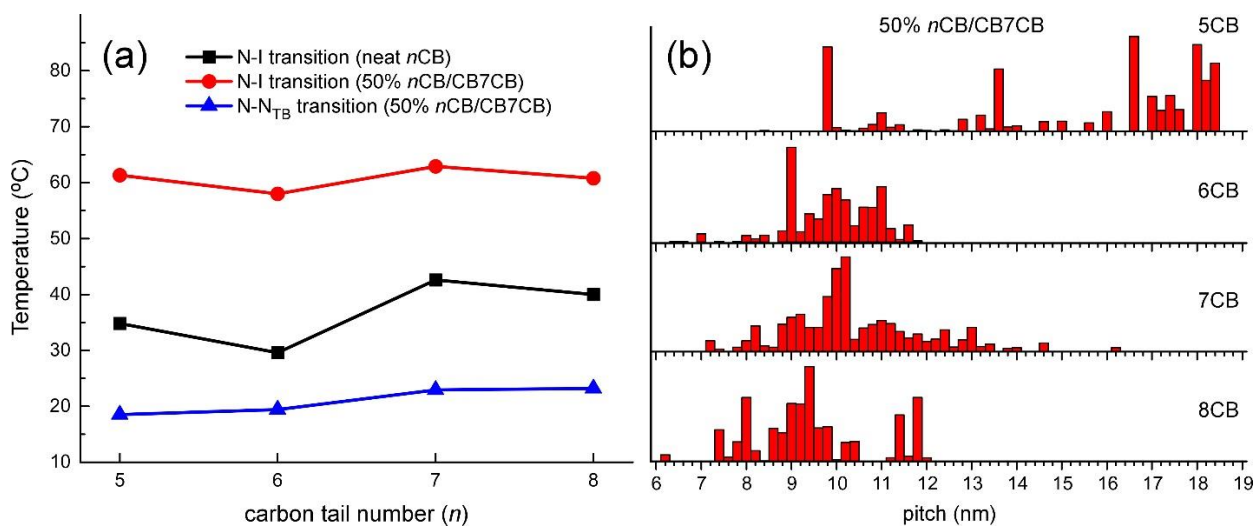

**Supplementary Figure S8. Transition temperatures and helix pitch distributions of the mixtures 50% *n*CB/CB7CB (*n* = 5 – 8) as measured by FFTEM.** (a) The transition temperatures of 50% *n*CB/CB7CB and neat *n*CB indicate that the odd-even effect in the Iso–N transition temperature of the *n*CB series is also present in the Iso–N transitions of the mixtures. Because we were unable to resolve the N–$N_{TB}$ transition temperature in these mixtures by DSC, we recorded the onset of each phase transition using PLM. (b) Measured helix pitch distribution as a function of carbon tail number *n* in 50% *n*CB/CB7CB mixtures. The pitch distributions of the mixtures are quite broad, and show no clear dependence on *n* except for a very subtle decrease in the pitch with increasing *n*.

| sample | N–N$_{TB}$ transition temperature (°C) | quench temperature (°C) | # of FFTEM images | # of pitch measurements |
|---|---|---|---|---|
| Neat CB7CB | 100.4 | 100 | 56 | 144 |
| | | 95 | 66 | 143 |
| | | 90 | 79 | 123 |
| | | 25 | 62 | 95 |
| | | | | |
| 12.5% 5CB/CB7CB | 81.4 | 68 | 28 | 33 |
| 25% 5CB/CB7CB | 61.4 | 63 | 30 | 38 |
| | | 58 | 33 | 43 |
| | | 50 | 34 | 58 |
| | | 38 | 30 | 74 |
| | | 29 | 21 | 47 |
| | | 25 | 99 | 89 |
| 37.5% 5CB/CB7CB | 41.4 | 25 | 48 | 71 |
| 50% 5CB/CB7CB | 19.9 | 8 | 65 | 59 |
| 62.5% 5CB/CB7CB | -0.9 | -17 | 62 | 85 |
| | | | | |
| 50% 6CB/CB7CB | 20.8 | 8 | 54 | 72 |
| 50% 7CB/CB7CB | 24.6 | 8 | 94 | 190 |
| 50% 8CB/CB7CB | 27.5 | 8 | 41 | 81 |

**Supplementary Table 1. Phase transition temperatures, FFTEM quench temperatures, and pitch analysis information for CB7CB and nCB/CB7CB mixtures.**

**Supplementary Discussions 1: Chemical synthesis of CB7CB**

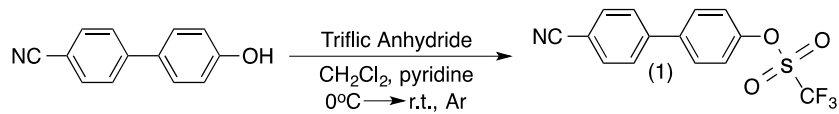

**1,1,1-trifluoro Methanesulfonic acid-4'-cyano[1,1'-biphenyl]-4-yl ester (1)**

5 g (25.6 mmol) 4-cyano-4'-hydroxybiphenyl was dissolved in 85 mL of $CH_2Cl_2$, cooled to 0°C and placed under argon. 32 mL (32.0 mmol) of 1 M triflic anhydride in $CH_2Cl_2$ was added drop wise 3.5 mL (43.5 mmol) of anhydrous pyridine was then added via syringe and reaction warmed to room temperature overnight. After 20 hours, the reaction mixture was poured over ice, diluted with $CH_2Cl_2$, washed once with water, once with 3% $H_2SO_4$, once with saturated aqueous NaCl, dried over $MgSO_4$, and reduced pressure to afford 8.11 g (24.8 mmol, 97% crude yield) light orange solid.

$^1$H NMR (400 MHz, $CDCl_3$): 7.80-7.72 (m, 2H), 7.71-7.62 (m, 4H), 7.45-7.36 (m, 2H).

$^{13}$C NMR (101 MHz, "$CDCl_3$"): 149.69, 143.61, 139.58, 132.78, 129.10, 127.83, 122.08, 118.71 (q, $CF_3$), 118.53, 111.85, 77.31, 76.99, 76.83, 76.67.

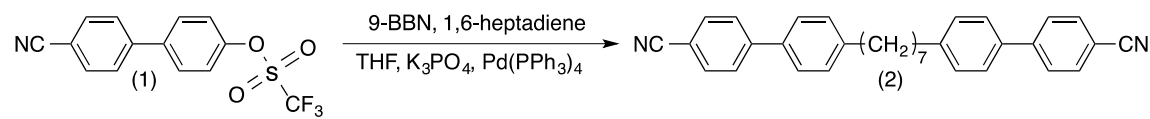

**4',4'-(heptane-1,7-diyl)bis(([1',1''-biphenyl]-4''-carbonitrile)) (2)**

1.18 g (12.2 mmol) 1,6-heptadiene was dissolved in 20 mL dry THF and placed under argon. 51 mL (26.0 mmol) 0.5 M 9-Borabicyclo(3.3.1)nonane (9-BBN) in THF was added and the mixture was heated and refluxed for 3 hours. To a second flask 7.85 g $K_3PO_4$ (37.0 mmol), 6.89 g (**1**) (21.1 mmol), and 40 mL THF were added and sparged with argon for 1 hour. 815 mg $Pd(PPh_3)_4$ (.710 mmol) dissolved in 20 mL THF was then cannulated into the second flask, followed by the 9-BBN adduct from flask 1. The contents of flask two were refluxed for 3 days under argon. THF was removed under reduced pressure and the crude oil was subsequently dissolved in $CH_2Cl_2$. The solution was washed twice with water, once with saturated aqueous NaCl, dried over $MgSO_4$, and placed under reduced pressure to afford a dark brown oil. The crude oil was purified via flash chromatography with an eluent gradient of 95:5:2 hexanes:ethyl acetate:$CH_2Cl_2$ to 80:20:2 hexanes:ethyl acetate:$CH_2Cl_2$, then recrystallized multiple times in hexanes and acetonitrile to afford 2.11 g (4.64 mmol, 38% yield) white solid.

$^1$H NMR (300 MHz, $CD_2Cl_2$): 7.77-7.64 (m, 7H), 7.59-7.48 (m, 4H), 7.35-7.24 (m, 4H), 2.71-2.60 (m, 4H), 1.65 (q, $J$ = 7.4 Hz, 4H), 1.42-1.30 (m, 6H).

$^{13}$C NMR (75 MHz, $CDCl_3$): 145.69, 143.80, 136.61, 132.68, 129.29, 127.58, 127.20, 119.15, 110.68, 35.73, 31.48, 29.47, 29.33.